\documentclass[twocolumn,showpacs,preprintnumbers,amsmath,amssymb]{revtex4}
\usepackage{amsmath,amsfonts,amssymb}
\usepackage{epsfig}
\def\bit{\begin{itemize}}
\def\eit{\end{itemize}}
\def\ben{\begin{enumerate}}
\def\een{\end{enumerate}}
\def\bed{\begin{description}}
\def\eed{\end{description}}

\def\lsim{\raise0.3ex\hbox{$<$\kern-0.75em\raise-1.1ex\hbox{$\sim$}}}
\def\gsim{\raise0.3ex\hbox{$>$\kern-0.75em\raise-1.1ex\hbox{$\sim$}}}

\let\jnfont=\rm
\def\NPB#1,{{\jnfont Nucl.\ Phys.\ B }{\bf #1},}
\def\PLB#1,{{\jnfont Phys.\ Lett.\ B }{\bf #1},}
\def\EPJC#1,{{\jnfont Eur.\ Phys.\ Jour.\ C }{\bf #1},}
\def\PRD#1,{{\jnfont Phys.\ Rev.\ D }{\bf #1},}
\def\PRL#1,{{\jnfont Phys.\ Rev.\ Lett.\ }{\bf #1},}
\def\MPLA#1,{{\jnfont Mod.\ Phys.\ Lett.\ A }{\bf #1},}
\def\JPG#1,{{\jnfont J.\ Phys.\ G}{\bf #1},}
\def\CTP#1,{{\jnfont Commun.\ Theor.\ Phys.\ }{\bf #1},}
\def\JHEP#1,{{\jnfont JHEP \ }{\bf #1},}
\def\NPPS#1,{{\jnfont Nucl.\ Phys.\ Proc.\ Suppl.\ }{\bf #1},}

\def\beq{\begin{equation}}
\def\eeq{\end{equation}}
\def\bea{\begin{eqnarray}}
\def\eea{\end{eqnarray}}
\newcommand{\ba}{\begin{array}}
\newcommand{\ea}{\end{array}}

\begin{document}
\title{Analysis and experimental study on the Jumping Chain}

\author{Wenyu Wang, Wu-Long Xu, Yang Xu,  Xu-Dong Yang}

\affiliation{Faculty of Science, Beijing University of Technology, Beijing, China}

\begin{abstract}
A freely falling chain from a cup at certain height can jump.
The process can be divided into two parts: a stable suspension 
 and an accelerating procedure.
Variational principle and force analysis demonstrate that the
shape of stable suspension is
an inverted catenary. The requirement of the jumping and 
the parameters to describe the jumping
catenary have been studied in detail, and experiments have been conducted
to verify the theoretical analysis.
The physical picture  
of the falling chain could be useful in certain falling systems,
 providing valuable insight into the dynamical system. 
\end{abstract}
\maketitle
\section{Introduction}\label{sec1}
\begin{figure}
\begin{center}
\scalebox{0.38}{\epsfig{file=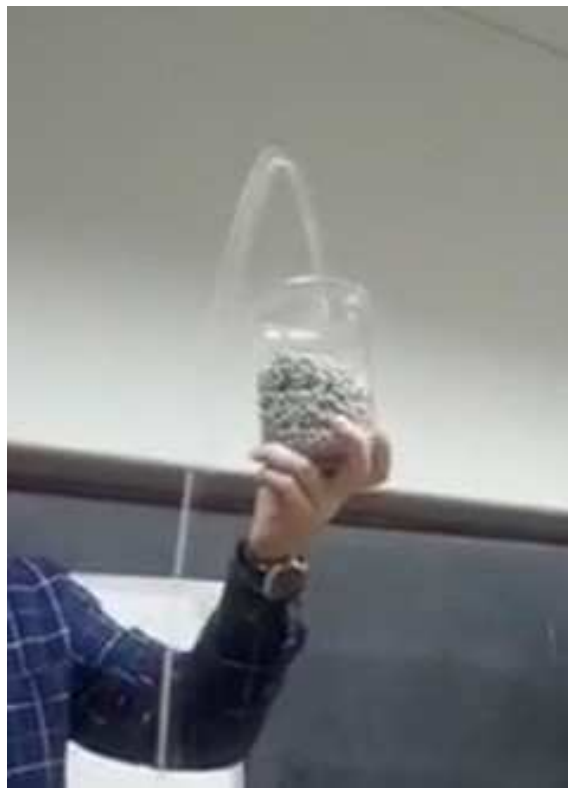}}
\caption{The jumping chain.}
\label{fig1}
\end{center}
\end{figure}
There is a very interesting physical phenomenon 
as shown in FIG. \ref{fig1}: a coiled but not entangled chain is put 
in a cup. When the end of the chain is pulled out of the cup,
it will fall freely along the wall of the cup. 
At a certain point, usually in a sudden moment, the chain can jump up 
and remain in a steady arc shape which can last for a certain
duration in the air. This experiment is often demonstrated  in college 
physics classes as an example  of an 
intriguing classical mechanical problem.
The reader can also observe the details of the jumping in the 
supplementary video or on the internet website~\cite{vedio}.
In this paper, we attempt to analyze and study the pattern and
related topics such as the height of the jump and the height 
of the fall, {\it etc.}  Experiments will be conducted and 
the data will be analyzed to verify our theoretical analysis~\cite{wxxy}.

In fact, the phenomenon is an application of classical 
mechanics in string, rope or fluid physics~\cite{jordan}. The jumping rope is a familiar example of a slender structure that interacts
with the fluid through which it moves~\cite{aerodynamics}.
The chain is usually  suspended and swung by a mechanical apparatus,  
and the pattern is well studied in the literature~\cite{hanging}.
This is because important class of truly 1D problems involves the motion of lines or filaments embedded in 3D space, such as vortex lines in
a defect lines in liquid crystals, type-II superconductor, 
or polymers in imposed flows.  Nonlinear partial differential 
equations describes the dynamics, and the twisting, breaking, reconnecting, or the knotting,  are the 
essential topological parts of the chain dynamics~\cite{topology}.
However, the jumping chain studied in this paper, 
is not suspended but jump from a higher platform, falling
freely in the space under the attraction of gravity. 
The damping from the viscosity of 
the air seems to be a sub-dominant effect in the whole process.
The gravity and the tension inside the chain play an important role.
The subtle point is that, although the 
matter is moving from the higher position to the lower
ground, the system is not a fluid or a falling stream. A detailed
study will reveal the different dynamics. 

In all, the paper is organized as follows: 
The phenomenology and dynamics of the jumping chain 
will be analyzed in Sec.~\ref{sec2}; The
experimental verification of  our analysis will be presented
in Sec.~\ref{sec3};
The conclusion is given in Sec. \ref{sec4}.

\section{Phenomenology and Theoretical Analysis}\label{sec2}
\subsection{Phenomena}
First, let's  describe  how the jumping phenomenon takes place.
As shown in FIG.~\ref{fig1}, the chains are coiled layer 
by layer inside a cup, and it is kept from becoming  entangled
 (It is difficult to keep all steel balls untangled
in the preparation, and it is one of the dominant disturbing
factors  of the experiment.). 
Then the cup is placed at a certain height.  
a small length of chain is taken out of the cup 
and allowed to  fall freely.
In the actual operation,  we also 
tried to use a soft and non-contractible rope to do the same experiment. 
The rope can barely jump up but the phenomenon is rather obscure,
so the thin metal chain composed of steel balls is adopted
in our experiment. The steel balls are loosely  connected, and the 
chain is  non-retractable. It 
means that the  chain do not have flexural capacity. 
This is the key difference between the chain and the soft rope.
The reason will be discussed in the following context.

The experiments show that the chain jumps
to appropriate higher position from the cup if the cup is set at 
some specific heights.
Due to the complex way of chain winding in the cup, 
the starting point of the jump keeps  changing, so the
shape of the jumping from cup to the highest point is very complicated. 
As for the pattern between the highest point and the falling off point
on the ground, the chain can maintain an almost stable shape 
near the highest point on the top.
The shape keeps  shaking near the ground, but this 
does not affect the stable shape on the top.

\begin{figure}
\begin{center}
\scalebox{1.2}{\epsfig{file=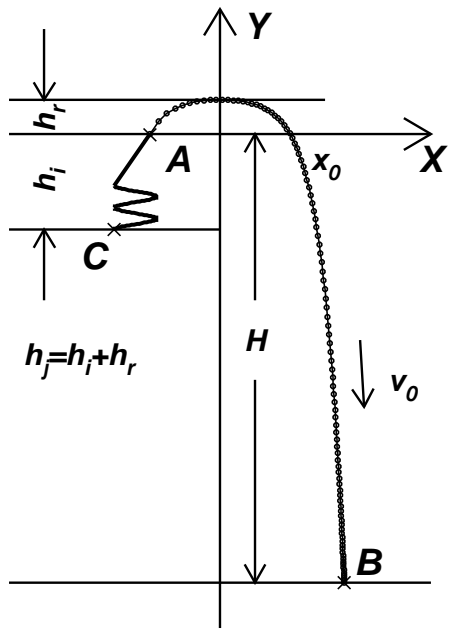}}
\caption{The sketch map of our analysis on the jumping chain.}
\label{fig2}
\end{center}
\end{figure}

Summarizing our observations, we think this physical phenomenon
can be divided into two sub-processes: the accelerating
process and the stable suspension process.
Thus our analysis on the phenomenon is set out as 
shown in FIG.~\ref{fig2}, in which point A is the beginning
of the suspension and the end is point B, which 
is the collision point on the ground.
The point C is the starting point of the jumping
in the cup where the elements of the chain are static.
The part from C to A is the accelerating process.
Note that, as discussed above, point B keeps shaking,
and point C also varies quickly.  At the same time,
the height of Point C is decreasing. The definitions of 
the variables such as $h_j,~h_i$ {\it etc.} are given
in the following context. One can see that
the stable suspension is obviously of the most concern,
and according to following analysis, we find that this is
actually the key point to comprehend the phenomenon.
Thus we analyze  the stable suspension in the following subsection.

\subsection{Analysis on the shape of the stable Suspension}
At first glance, the jumping chain appears like a parabolic curve.
However, the truth is not so simple as expect.
A free-falling mass point follows a parabolic trajectory.
However, for the non-retractable chain,
each element on the suspended chain falls 
at the same velocity, only the directions vary.
Based on our observations, the coiled chain in the cup
almost moves to the ground at a constant rate. thus it is
natural to assume that the suspended chain can maintain
a constant velocity $v_0$ for appropriate interval as 
an ideal case. Then, in $\Delta t$ time, there is a part of the
chain with mass $\lambda v_0dt$  ($\lambda$ is denoted as 
the line density) in the cup which is accelerated from 0 to $v_0$, 
At the point A, the chain feels the dragging tension from 
the chain at rest in the cup as follows:
\begin{equation}
T_0=\lim_{\Delta t \to 0}
 \frac{\lambda v_0^2 \Delta t}{\Delta t}=\lambda v_0^2\, .
\end{equation}
It should be noted that this is merely the ideal case, and the following study will show that the actual tension   
at the end of A will be less than $\lambda v_0^2$. 
At point B, the chain element with velocity $v_0$ collides with the 
ground. If the collision is elastic, the chain will rebound from the
ground with velocity $v_0$. Similar analysis gives that 
impulsive force to the ground is $2\lambda v_0^2$.
Of course, the collision is not completely elastic due to the 
dragging. We shall provide the specific method for dealing with the 
impulsive force at point B later.

We find that, very interestingly, the shape of the suspended chain 
is actually an inverted catenary. The exploration of the catenary is an 
important progress in the history of classical mechanics and mathematics.
Our discovery in fact compensates a block in the catenary theory. 
Here we briefly show the derivation of the catenary and 
explain the reason for the same shape of the suspended chain.
Suppose the shape of the chain is $y=y(x)$, the solution of the
catenary is to find a $y(x)$ such that the potential
or center of gravity of the chain is at its lowest point.
Using the variational principle, the action of the whole system is
the center of gravity in the vertical direction
\begin{equation}
  S=-\int \lambda g(y-c)\sqrt{1+{y^\prime}^{2}}dx=\int\mathcal{L}dx,
\end{equation}
where $g$ is the acceleration of gravity and $c$ is an arbitrary
constant, $y^\prime$ is the  derivative of $y$. $\mathcal{L}$ 
is the Lagrangian of the system
\begin{eqnarray}
&&\mathcal{L} = -\lambda g(y-c)\sqrt{1+{y^\prime}^{2}}\label{motion2}.
\end{eqnarray}
The motion for $\mathcal{L}$ gives the non-linear differential equation of
the catenary
\begin{equation}
 (y-c)y^{\prime\prime}-{y^\prime}^2-1=0.\label{motion3}
\end{equation}
The solution is
\begin{equation}
  y=c+\frac{a}{2}(e^{\frac{x}{a}+b}+e^{-\frac{x}{a}-b}).
\end{equation}
where $a,~b$ are the parameters determined by the inputs of the catenary.
The approach described above is the standard derivation of the catenary. 
Note that here suspension chain are hanged at any height,
thus an additional constant $c$ are introduced.
In fact, we notice that, although the equation of motion 
Eq.~\eqref{motion3} is nonlinear, it has a discrete symmetry and $y(x)$ 
can be transformed
\begin{equation}
  y(x)\rightarrow Y(x)=-y(x)+2c.
\end{equation}
$Y(x)$ is also a solution of the catenary equation. 
Despite its lack of practical application,  the inverted
catenary solution is still a valid solution of the catenary equation.
This solution is an unstable extremum
point of variation, which has been largely overlooked in the literature.
Nevertheless, our research has revealed that this inverted solution is 
indeed the exact shape of the suspended chain.

For the jumping chain, the potential energy is the same as 
that of the catenary. Therefore only the kinetic term 
needs to be included in the action
\begin{equation}
  S=\int \mathcal{L}dx=\int \left(\frac{1}{2}\lambda\sqrt{1+{y^\prime}^2}v_0^2
-\lambda gy\sqrt{1+{y^\prime}^2}\right)dx\, .
\end{equation}
 By comparing Eq.~\eqref{motion2} and assuming that $v_0^2$ is a constant, 
 we can observe that the Lagrangian of the jumping chain
 is identical to that of the  catenary,
 with the only difference being that  $c$ is taken as a constant
\begin{equation}
  c=\frac{v_0^2}{2g}\, .
\end{equation}
Due to the given conditions, the solution will be
a convexity, specifically, an inverted  catenary
\begin{equation}
  y=c-\frac{a}{2}(e^{\frac{x}{a}}+e^{-\frac{x}{a}})\,.
\end{equation}

\begin{figure}
\begin{center}
\scalebox{0.42}{\epsfig{file=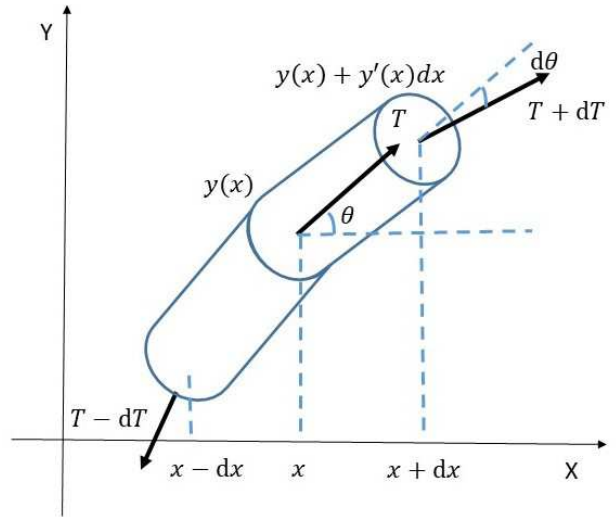}}
\caption{The force analysis on the elements of the jumping chain.}
\label{fig3}
\end{center}
\end{figure}

The difference between the deviation of an ordinary catenary and 
a jumping chain is that the chain is constantly  
in motion. The elements of the chain is continually jumping out of the 
cup and falling on to the ground, thus changing the mass of 
system. This may lead some readers to be skeptical about the validity
of variational principle. However, the same differential 
equation of motion can be derived by the force analysis.
As shown in FIG.~\ref{fig3}, 
the tension in the jumping chain is denoted as $T=T(x)$
and the convention of the forward direction is defined 
as from point A to point B. Every element of the chain
is moving in a curvilinear motion with the same speed, so
the tangential acceleration is zero, and the infinitesimal
tension $dT$ cancels the tangential component of gravity
\begin{eqnarray}
    dT=\lambda g\sqrt{1+{y^\prime}^2}dx  \frac{y^\prime}{\sqrt{1+{y^\prime}^2}}
    =\lambda g\frac{dy}{dx} dx=\lambda g dy\,.
\end{eqnarray}
Thus the tension in the chain can be expressed as
\begin{equation}
	T=T_A+\lambda gy\,.\label{zhangli}
\end{equation}
This is a very interesting result that 
the tension in the stable jumping chain changes along with the height of
the elements, which is  the same as the tension in the
ordinary catenary.
To analyze the normal component force of the infinitesimal 
chain imposed by tension from the curvature of the chain,
we can look to the  catenary or hanged rope for comparison.
In the case of the ordinary catenary,
 the normal component of the tension is canceled by the gravity,
 but in the case of the jumping chain,
the normal component force provides the centripetal force
together with the gravity. The geometrical configuration 
is shown in FIG.~\ref{fig3} where we
investigate the centripetal force acting on the two chain elements
$x-dx\to x$ and $x\to x+dx$. From this, we 
 can easily derive  the  equation for the
centripetal force 
\begin{eqnarray}
\frac{2\lambda \sqrt{1+y'^{2}}dx v^2}{\rho}= 
	 2T d\theta+2\lambda g \sqrt{1+y'^{2}} dx\frac{1}{\sqrt{1+y'^{2}}}\, ,
  \nonumber 
\end{eqnarray}
where $\rho$ is the radius of curvature at point $x$.
By substituting  $\rho$, $d\theta$, and tension $T$, we can get
\begin{eqnarray}
	\frac{2\lambda|y''|v_0^2 dx}{1+y'^2}=\frac{2(T_A+\lambda g y)|y''|dx}{1+y'^2}+2\lambda g dx\,. \label{ten1}
\end{eqnarray}
Due to the convexity of the jumping chain
\begin{eqnarray}
	|y^{\prime\prime}|=-y^{\prime\prime}\, .
\end{eqnarray}
Eq.~\eqref{ten1} are reduced to 
\begin{eqnarray}
-(T_A+\lambda gy -\lambda v_0^2)y''+\lambda g y'^2+\lambda g=0\, .
\end{eqnarray}
Given that $T_A$ is the variable to be solved, it is convenient
to define
\begin{eqnarray}
	T_A-\lambda v_0^2=-\lambda gc \label{tac}\, .
\end{eqnarray}
Then the differential equation of $y$ is
\begin{eqnarray}
	(y-c)y^{\prime\prime}-{y^{\prime}}^2-1=0\, .\label{def-cat}
\end{eqnarray}
Proved the same catenary equation again.

Although the analysis of force is a bit more complicate  
than that of the variational principle of mechanics, 
the physical picture is easier to understand because
it is an analysis of forces.
We can also see that the forces at the two ends of the catenary
and the jumping chain  are different.
In case of the ordinary catenary, the tensions at the two ends
are dragging the chain. Whereas, in the 
case of the jumping chain, the tension
at point A is dragging the chain, while the tension at 
point B is pushing the chain. Of course, this is the ideal case;
the jumping process will be further analyzed in the following.

\subsection{Analysis on the jumping}
First, let's begin with the explanation for why $T_A\neq\lambda v_0^2$.
$T_A=\lambda v_0^2$ corresponds to the ideal case  that 
each chain element instantaneously accelerates from 0 to $v_0$.
In contrast, the actual situation is that the chain is loosely coiled and 
the chain element needs an interval for the  acceleration,
making $T_A$ smaller than $\lambda v_0^2 $.
Analytically,  if $c=0$, the shape of the suspended chain line is
\begin{eqnarray}
	y=-\frac{a}{2}(e^{\frac{x}{a}}+e^{-\frac{x}{a}})\,,
\end{eqnarray}
then 
\begin{eqnarray}
	y\le -a\,.
\end{eqnarray}
It is evident that  $y>0$ solution does not exist, as indicated by 
Eq.~\eqref{def-cat}, which a non-linear differential equation.
Therefore, if the coordinate system shown in FIG.~\ref{fig2} is used
a $y>0$ solution requires a non-zero $c$. 
From Eq.~\eqref{tac}, it is clear that
$T_A$ must be less than $\lambda v_0^2$.
As shown in Fig.~\ref{fig2}, the acceleration process
for the upper elements in the stable suspended shape is added.
And the height of the acceleration jumping up process 
from point C to point A is measured as $h_i$.
The jumping height of the stable suspended chain line is $h_r$.
The total jumping height is thus the sum of $h_i$ and $h_r$
\begin{eqnarray}
	h_j = h_i +h_r\, .
\end{eqnarray}
Along with the conventions mentioned above, 
the falling height (the height of the cup) of the chain is denoted as $H$ 
and the width of the jump is $2x_0$ as shown in Fig.~\ref{fig2}.
We will  further investigate the relationship between these 
quantities for jumping chain and attempt to
conduct  experiment to verify our analysis.

The tension at point A is
\begin{eqnarray}
	T_A=\lambda v_0^2-\lambda gc\,. 
\end{eqnarray}
According to the Eq.~(\ref{zhangli}), 
the tension $T$ from point A to point B is
\begin{eqnarray}
	T=\lambda v_0^2+\lambda gy -\lambda gc\,. 
\end{eqnarray}
As discussed previously,  the elastic collision implies that 
the force imposed from the ground to the chain at point
B will be $2\lambda v_0^2$. This may appear suspicious to some
readers, as a deceleration process is needed at point B.
When the chain finally fall on the ground, the tension should
be less than $2\lambda v_0^2$. However, 
we found that the chain elements can slightly bounce on the ground, 
thus the actual impulsion to the ground should be greater 
than $\lambda v_0^2$. To account for the force with more precision,  
a constant coefficient $\alpha$ can be introduced
\begin{eqnarray}
  -\alpha\lambda v_0^2=\lambda v_0^2-\lambda gH-\lambda gc\, .
\end{eqnarray}
The value of $\alpha$ would be greater than 1 and less than 2. 
It should be noted that 
tension at point B should be negative according the 
direction defined above. Since falling height $H$ is much larger
than the jumping height $h_j$,  the tension fluctuation 
at point B is not sensitive to stable suspension above. Then
\begin{eqnarray}
  v_0^2=\frac{1}{\alpha+1}g(H+c)\label{v0}\,. 
\end{eqnarray}
This is the relation between the falling height and the velocity
of the suspended chain. 
The jumping height $y(x=0)$ of the stable suspended chain is 
\begin{eqnarray}
  h_r=c-a\,. 
\end{eqnarray}
Then all the exploration  focuses on the pending
parameter $c$ and $\alpha$, which will be analyzed in the
next section when discussing the experimental verification.
Before that, let us analyze the acceleration process.

As the velocity of the chain element changes over time during
the acceleration process, it seems difficult to do the analysis. However,
the time parameter can be eliminated after some substitutions.
The velocity of accelerating elements is denoted as $v$. According to
Fig.~\ref{fig2} and Eq.~\eqref{zhangli}, the tangential acceleration
should be 
\begin{eqnarray}
  \lambda vdt\frac{dv}{dt}=\lambda v dv=dT-\lambda gdy\,. \label{jia}
\end{eqnarray}
Since the chain elements are loosely connected, the normal
acceleration dynamical equation can be disregarded. It can be seen that
the time dependence $dt$ can be eliminated. This is the 
mathematical formulation of the transformation from gravity potential to 
kinetic energy. Carry out the  integration on both sides, we have 
\begin{eqnarray}
  \int ^{v_0}_{0}\lambda v dv&=&\int_0^{T_A} dT-\int_{-h_i}^0\lambda gdy\nonumber\\
&=&T_A-\lambda gh_i\,.
\end{eqnarray}
At point A
\begin{eqnarray}
  \frac{1}{2}\lambda v_0^2 =\lambda v_0^2 -\lambda g c -\lambda  g h_i\,,\nonumber
\end{eqnarray}
namely
\begin{eqnarray}
  h_i = \frac{v_0^2}{2g} -c\,.
\end{eqnarray}
Substitute $v_0$ from Eq.~\eqref{v0}, we can get the 
relation between $h_i$ and $H$
\begin{eqnarray}
  h_i =\frac{H-(2\alpha +1)c}{2(\alpha+1)}\,.
\end{eqnarray}
From above equation,
the requirement for the jumping can also be got. The jump
can happen only when $h_i$ greater than zero, thus the condition 
for the jumping is 
\begin{eqnarray}
  H>(2\alpha+1) c \label{qitiao}\,.
\end{eqnarray}
The total jumping height is
\begin{eqnarray}
 h_j =h_i+h_r =\frac{H+c}{2(\alpha+1)}-a\label{hjh}\,.
\end{eqnarray}
It is evident that the jumping height basically is linearly correlated
with the falling height $H$, with
additional parameter $c$, $a$ and $\alpha$ to be determined.

\section{Experimentation and Verification}\label{sec3}

\subsection{ Analysis and execution of the experiment}\label{sec3.1}

\begin{figure}
\begin{center}
\scalebox{0.42}{\epsfig{file=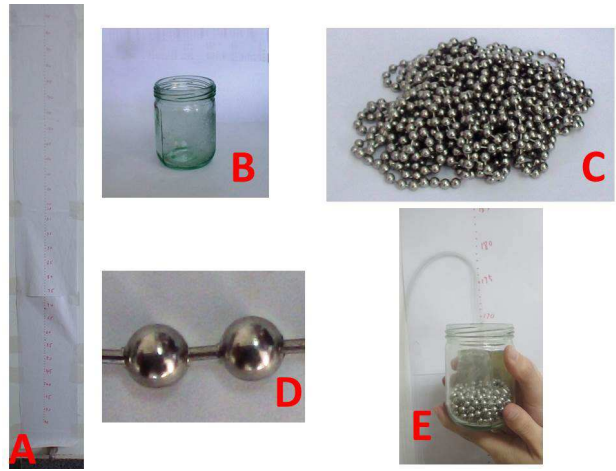}}
\caption{The apparatus of the experiment: A, the height 
calibration on the wall;
B, the cup; C, the chain; D, the chain elements, E, the screenshot.}
\label{fig4}
\end{center}
\end{figure}

Through the theoretical analysis in Sec.~\ref{sec2}, we can verify
the relation through experimentation. Nevertheless, it is important
to bear in mind that all the theoretical analysis are
based on ideal conditions, and  there are certain
limitations in the actual experiment, such as
\begin{enumerate}
    \item Ideally, the chain in the cup should be loosely
    connected and freely coiled. Unfortunately, the elements in chain
    inevitably become  tangled, which prevents the chain from 
    maintaining homogeneous during the the acceleration 
    and stable suspension
    processes. We discovered that this is the primary source of 
    disturbance in the experiment. Even a small knot can cause 
    a significant variation of the stable suspension.
    \item Ideally, the jumping point C and falling point B should remain
     fixed in order to obtain a  stable suspension. However, these 
     two points tend to  shake during the falling process,
     making it difficult to accurately measure the shape. 
     Nevertheless, Since the shape of catenary is an exponential function
     in which $y$ increases exponentially with $x$, the shaking chain
     almost forms a straight line at the bottom. This makes it 
     challenging to measure the correlation between the shape and the exponential curve. Furthermore, due to the upper two points, it is difficult to precisely determine the 
     positions of points A, B and C. Generally, only the total
     jumping height $h_j$ and falling height $H$ can be measured. 
\end{enumerate}
Though there are these difficulties, it is sure that
\begin{eqnarray}
  H\gg c,~a\,.
\end{eqnarray}
and the jumping height $h_j$, $a$ and $c$ are at the same order. Based
on this, we can assume that the actual shape of the 
jumping chain resembles and  inverted catenary.
The experiment should then test the linear
relation between $h_j$ and $H$ (Eq.~\eqref{hjh}) and 
the linear relation between $v_0^2$ and $H$ (Eq.~\eqref{v0}). 
Both relations could be used to evaluate the parameter $\alpha$. 
which will validate our analysis in Sec.~\ref{sec2}. 

\begin{figure*}
\begin{center}
\scalebox{0.52}{\epsfig{file=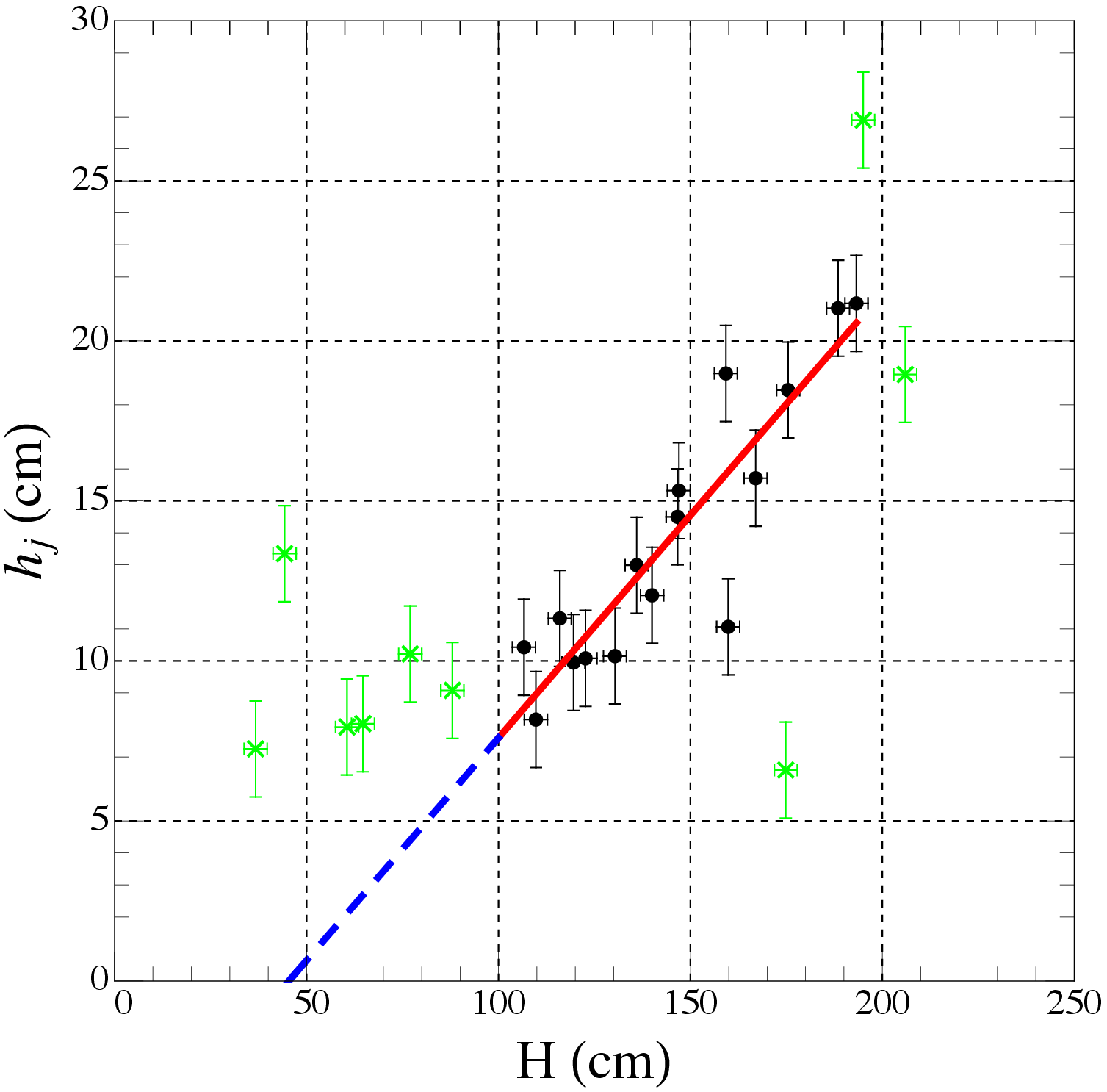}}
\scalebox{0.52}{\epsfig{file=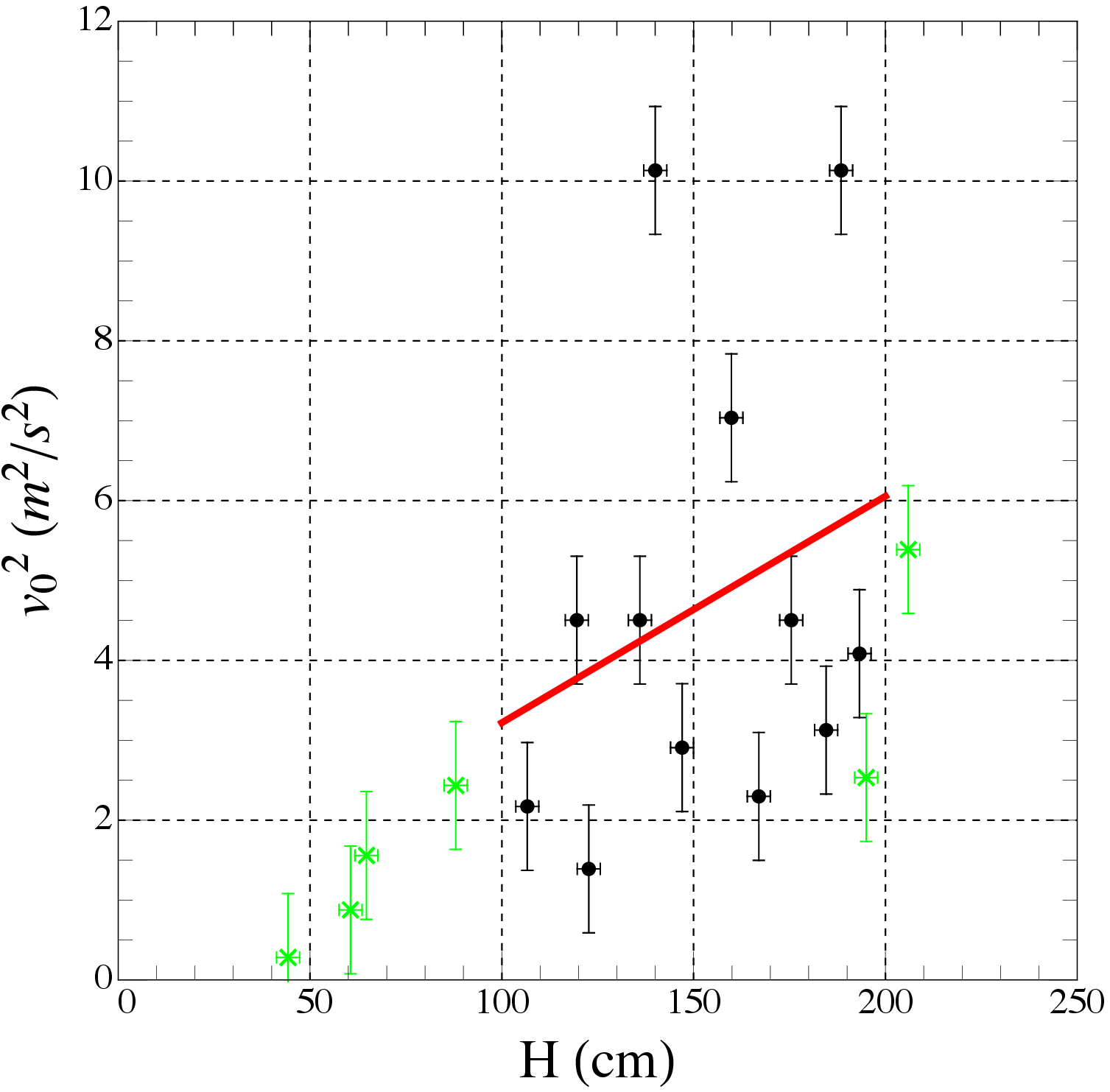}}
\caption{Left: The recorded results between the jumping height 
and falling height; Right:  the velocity $v_0^2$ of the stable
shape and the falling height}
\label{fig5}
\end{center}
\end{figure*}

The apparatus used in our experiment is illustrated
in Fig.~\ref{fig4}.
We drawn a height calibration on the wall,  which can be seen 
in penal A of Fig.~\ref{fig4}. The chain and the cup are also depicted
in the figure, and the details of chain are provided 
in the Tab.~\ref{tab1}.
  \begin{table}[htbp]
     \centering
     \begin{tabular}{cc}
     \hline
     \hline
   Diameter of the steal ball   & $5\pm0.5$mm\\
   Length of the steal ball & $3\pm0.5$mm\\
   Length of the chain     & 8.4$\pm$0.1m\\
   Density $\lambda$   & 0.036$\pm$0.03kg/m\\
   Height of the cup     & $95\pm0.5$mm\\
   Diameter of the cup     & $66\pm0.5$mm\\          
     \hline
     \hline
     \end{tabular}
     \caption{The setup of the exepriment.}
     \label{tab1}
 \end{table}
The experiment was conducted as follows: 
The chain was pulled out of the cup 
from a height 200cm, then gradually lowered by  5cm 
each time. When the height reached 100cm, it became 
increasingly 
difficult to lower the chain further. Therefore,
we chose to lower the chain by 10cm for heights lower than 
100cm, until the height reached 40cm. At each specific
height, a cellphone was used to record the jumping chain,
while the camera were placed at the same height at the  top of 
jumping chain (see penal E of Fig.~\ref{fig4}). 
Due to the presence of various disturbances 
in the experiment, multiple vedios were taken in order to select
the best one for the data collection.

 In Principle, catenaries with different $c$ and  $a$ can exist
 when the requirements discussed in Sec.~\ref{sec2} are satisfied.
 However, since the chain needs to jump from the cup, thus the shape
 of the jumping chain depends on specific initial conditions,
 namely, the diameter of the cup, the depth of the 
 cup at the beginning of the jump, {\it etc}. Thus though
 the shape of the jumping at a specific height 
 is not in a single parameters set,  We found
 that $c$, $a$ are at the same order of the diameter of the cup.
 thus the linear relation could be found in the following. 

The data collection was carried out as follows. each frame
of the chosen vedio was examined, and the best moment with the highest
jumping height and stable shape was identified. The falling height
and jumping height were then recorded base on the frame at the best 
moment. To measure $v_0$, the 
time interval between a specific height of the chain left in the cup
and the final moment was recorded. With these measured raw data, we can 
analyze on the relation between the physical quantities talked 
in the Sec.~\ref{sec2}

\subsection{Analysis on the data}\label{sec3.2}
As discussed in the previous subseciton, linear
relation between $h_j$ and $H$ (Eq.~\eqref{hjh}) and 
the linear relation between $v_0^2$ and $H$ (Eq.~\eqref{v0})
can be used to verify  our analysis.
The corresponding results are shown in Fig.~\ref{fig5},
with the left panel displaying the recorded jumping heights
and falling height, and the right panel shows the calculated
velocity $v_0$ of the stable suspension shape and falling height. Note
that due to presence of various disturbances in the experiment, 
only data with falling higher than 100cm (black points) were chosen for 
the analysis. As can be seen in Fig.~\ref{fig5}, the linear relations 
are realized and the slopes give  
the values of the $\alpha_1$ (left panel) and $\alpha_2$ (right panel)
\begin{eqnarray}
    \alpha_1 = 2.6^{+5}_{-1.3},~~~ \alpha_2 = 2.5^{+7.0}_{-0.7}\,. 
\end{eqnarray}
We can see that the errors are quite large and the central values
exceed 2. However, given the difficult setup of the experiment,
these two measurements on $\alpha$ are acceptable.
Note that although $\alpha_1$ and $\alpha_2$ come from the same
series of derivations, Eq.~\eqref{hjh} and Eq.~\eqref{v0} are assumed 
to independent with each other. Thus,  we can conclude that 
$\alpha_1$ and $\alpha_2$ are in agreement with each other. 
Another interest point is that the extension line in the left panel
intersects the $H$ coordinate axis at a point greater than 0.
This point should correspond to  the required height of the jumping
condition in Eq.~\eqref{qitiao},
Although  this is only a rough estimation. 

Finally, we can explain  why the soft rope
cannot effectively jump. the elements of chain are connected loosely,
and the rope cannot achieve this condition. This is also
the reason why the catenary cannot be realized by the soft rope.
Specifically, the soft rope lacks the non-flexural capacity 
and the absence of disturbance from normal
force, making the jumping impossible.

\section{Conclusion}\label{sec4}
The jumping of a free-falling chain is an intriguing  phenomenon,
and this paper provides a  detailed analysis of  the physics
behind it. The analysis
are divided into two parts: the stable suspension and the jumping.
Both variational principle and force analysis demonstrate that  the 
stable suspension is an inverted catenary. The 
parameters that describe the phenomenon 
are studied and tested through experiments, The requirement for
the jumping are discussed. 
The measurements of the parameter $\alpha$ meet  our expectations.
The physical picture  of the falling chain could
be useful in some falling systems.
and the inverted catenary may be a complement to classical mechanics.

\end{document}